\documentclass[preprint,superscriptaddress,nofootinbib,dvipdfmx]{revtex4}
\usepackage{graphicx}
\usepackage{amsmath,amssymb,color}

\newcommand{\red}[1]{{\color[rgb]{1,0,0} #1}}

\newcommand{\MNS}{{\text{MNS}}}

\newcommand{\eV}{{\text{eV}}}

\newcommand{\GeV}{{\text{GeV}}}

\newcommand{\Det}{\text{Det}}

\newcommand{\U}{{\text{U}}}
\newcommand{\SU}{{\text{SU}}}

\newcommand{\ellp}{\ell^\prime}

\begin{document}

\preprint{UT-HET 113}

\title{
Probing Models of Dirac Neutrino Masses
via the Flavor Structure of the Mass Matrix
}

\author{Shinya Kanemura}
\email{kanemu@sci.u-toyama.ac.jp}
\affiliation{
Department of Physics,
University of Toyama,
3190 Gofuku,
Toyama 930-8555, Japan
}
\author{Kodai Sakurai}
\email{sakurai@jodo.sci.u-toyama.ac.jp}
\affiliation{
Department of Physics,
University of Toyama,
3190 Gofuku,
Toyama 930-8555, Japan
}
\author{Hiroaki Sugiyama}
\email{sugiyama@sci.u-toyama.ac.jp}
\affiliation{
Department of Physics,
University of Toyama,
3190 Gofuku,
Toyama 930-8555, Japan
}


\begin{abstract}
 We classify models of the Dirac neutrino mass
by concentrating on flavor structures of the mass matrix.
 The advantage of our classification is
that we do not need to specify
detail of models except for Yukawa interactions
because flavor structures can be
given only by products of Yukawa matrices.
 All possible Yukawa interactions between leptons%
~(including the right-handed neutrino)
are taken into account
by introducing appropriate scalar fields.
 We also take into account
the case with Yukawa interactions of leptons
with the dark matter candidate.
 Then,
we see that
flavor structures can be classified into seven groups.
 The result is useful
for the efficient test of models
of the neutrino mass.
 One of seven groups can be tested
by measuring the absolute neutrino mass.
 Other two can be tested
by probing the violation of the lepton universality
in $\ell \to \ellp \nu\overline{\nu}$.
 In order to test the other four groups,
we can rely on searches for new scalar particles
at collider experiments.

\end{abstract}

\maketitle

\section{Introduction}


 Discoveries of neutrino oscillations%
~\cite{solar, Aharmim:2011vm, Gando:2013nba,
Wendell:2010md, acc-disapp, Abe:2015awa,
s-reac, An:2015rpe, acc-app}
indicate that neutrinos have tiny but non-zero masses,
which is a clear evidence
for the new physics beyond the standard model~(SM).
 The SM must be extended to have neutrino masses.
 There are two possibilities for mass terms of $\nu_L^{}$,
which is the left-handed neutrino
in an $\SU(2)_L$-doublet lepton field
$L \equiv (\nu_L^{} \ \ell_L)^T$
with the left-handed charged lepton $\ell_L$.
 One is the Dirac mass term
$m_\text{D}^{} \bigl[ \overline{\nu_L^{}} \nu_R^{} \bigr]$,
for which right-handed neutrino $\nu_R^{}$ is introduced
as the singlet fermion under the SM gauge group.
 The other is the Majorana mass term
$(1/2) m_\text{M}^{} \bigl[ \overline{\nu_L^{}} (\nu_L^{})^c \bigr]$,
where the superscript $c$ denotes the charge conjugation.
 The Majorana mass term
violates the lepton number~(L\#) conservation by two units.
 If the Dirac mass term is generated via
the Yukawa interaction
$y_\nu [\overline{L} \epsilon \Phi^\ast \nu_R^{}]$
with the Higgs doublet field $\Phi$ in the SM,
where $\epsilon$ denotes $2\times 2$ antisymmetric matrix,
the Yukawa coupling constant $y_\nu$ must be
unnaturally small%
~($y_\nu \lesssim 10^{-12}$ for $m_\text{D}^{} \lesssim 0.1\,\eV$).
 On the other hand,
the Majorana mass term
is obtained from dimension-5 operators~\cite{Weinberg:1979sa},
e.g.\
$(1/\Lambda)
[\overline{L}\epsilon \Phi^\ast] [\Phi^\dagger \epsilon L^c]$,
where $\Lambda$ is the energy scale of the new physics.
 Then,
it seems to be an attractive feature of the Majorana neutrino mass
that the mass can be suppressed by a large $\Lambda$
without using extremely small coupling constants
as in the case of the seesaw mechanism~\cite{ref:seesaw}.



 Some of models of the neutrino mass
have common features.
 Classification of models according to such features
is useful for the efficient test of models
not one by one but group by group of them.
 The feature that is used for the classification
is desired to be model-independent
as much as possible.
 In Ref.~\cite{Kanemura:2015cca},
it was proposed to classify models
for Majorana neutrino masses
according to combinations of Yukawa matrices,
which give the flavor structure~(ratios of elements) of
the neutrino mass matrix
without specifying detail of models.
 In contrast,
the overall scale of the mass matrix
depends on details of models,
namely topologies~(tree level, one-loop level, etc.)
of Feynman diagrams for the mass matrix,
sizes of coupling constants in the diagram,
and masses of particles in the diagram.
 Classifications
according to topologies of diagrams~\cite{ref:diagram}
or higher-dimensional operators~\cite{ref:higher-dim}
are also useful to exhaust possible models.



 In Ref.~\cite{Kanemura:2015cca},
models that generate the Majorana neutrino mass matrix $m_\text{M}^{}$
were classified into three groups
according to combinations of Yukawa matrices.
 It was shown that
these groups can be tested
by measurements of the absolute neutrino mass%
~\cite{Osipowicz:2001sq, Abazajian:2013oma},
searches for $\tau \to \overline{\ell_1} \ell_2 \ell_3$%
~($\ell_1, \ell_2, \ell_3 = e, \mu$)~\cite{Abe:2010gxa},
searches for the neutrinoless double beta decay%
~($0\nu\beta\beta$. See e.g.\ Ref.~\cite{Dell'Oro:2016dbc}),
and neutrino oscillation experiments%
~(see e.g.\ \cite{Blennow:2013oma}).



 In this letter,
we classify models for the Dirac neutrino mass matrix $m_\text{D}^{}$
according to combinations of Yukawa matrices
subsequently to the work for the Majorana case in Ref.~\cite{Kanemura:2015cca}.
 The L\# conservation is respected
because the L\# violating phenomena such as $0\nu\beta\beta$
has not been observed so far.
 New physics models for the Dirac neutrino mass can be found in
e.g.\ Refs.%
~\cite{Ref:nuTHDM-D, Davidson:2009ha,
DSeesaw, 1loopDirac-LR,
1loopDirac, Kanemura:2011jj, Chen:2012baa,
Gu:2007ug,Farzan:2012sa,Okada:2014vla}%
~(see also Ref.~\cite{Babu:1989fg}).
  First, we do the classification
for models without new fermions except for $\nu_R$,
which has $\text{L\#}=1$.
 All possible Yukawa interactions between leptons
are taken into account
by introducing appropriate scalar fields.
 However,
we forbid $y_\nu [\overline{L} \epsilon \Phi^\ast \nu_R]$
because it requires unnaturally small $y_\nu$.
 Next,
we introduce $\psi_R^0$
as the singlet fermion under the SM gauge group with $\text{L\#}=0$
in order to have the dark matter candidate.
 We classify models
that have additional Yukawa interactions
of leptons with $\psi_R^0$,
for which scalar fields are further introduced.
 As the result of these analyses,
we find that these models
can be classified into seven groups.
 We also show how these groups can be
tested by $0\nu\beta\beta$ searches,
measurements of the absolute neutrino mass,
the lepton universality test
in $\ell \to \ellp \nu \overline{\nu}$,
and neutrino oscillation measurements
with/without additional information
from future collider experiments.


\section{Classification by Flavor Structures}


 In this section,
we classify models that generate Dirac neutrino masses
in order for efficient tests of them.
 For Dirac neutrino masses,
right-handed neutrinos $\nu_{i R}^{}$ with $\text{L\#} = 1$
must be introduced.
 The conservation of L\# is imposed,
which forbids Majorana mass terms
$(1/2) M_{iR} \Bigl[ \overline{(\nu_{iR}^{})^c} \nu_{iR}^{} \Bigr]$.
 The index $i$ runs from 1 to 3
in order to obtain three Dirac neutrino masses%
\footnote{
 If one of three neutrino is massless,
two $\nu_{iR}$ are enough.
}.
 If the Dirac neutrino mass
is generated via the tree level Yukawa interaction
$y_\nu [\overline{L} \epsilon \Phi^\ast \nu_R]$,
the Yukawa coupling constant $y_\nu$
must be unnaturally small.
 Even if we accept such a tiny coupling constant,
it makes the origin of the neutrino mass untestable.
 Therefore,
we assume that neutrino masses
are generated by a different mechanism.
 The tree level Yukawa interaction is forbidden
by introducing the softly-broken $Z_2$ symmetry%
~(we call it $Z_2^\prime$)
such that $\nu_R$ has the odd parity
while the SM particles have the even parity%
\footnote{
 Instead of the $Z_2^\prime$ symmetry,
we can impose the global $U(1)$ symmetry%
~(see e.g.\ Ref.~\cite{Davidson:2009ha}).
}.
 Then,
the Dirac neutrino masses can be generated
via the soft-breaking of the $Z_2^\prime$ symmetry.
 The soft-breaking parameters
are assumed to be in the scalar potential,
which we do not specify
in our model-independent analyses.



 Since we classify models
according to combinations of Yukawa matrices,
we must specify Yukawa matrices
that are used in our analyses.
 First,
we take into account
all possible Yukawa interactions
between leptons%
~(except for the tree level interaction
discussed in the previous paragraph).
 In order to have such interactions,
we introduce new scalar fields
as listed in Table~\ref{tab:particle-I}.
 Two scalar fields $s_R^+$ and $\Phi_\nu$ are
introduced as the $Z_2^\prime$-odd ones
so that they can provide Yukawa interactions
between $\nu_R^{}$ and leptons.
 Although
we forbid $y_\nu [\overline{L} \epsilon \Phi^\ast \nu_R]$,
the Yukawa interaction
$Y_\nu [\overline{L} \epsilon \Phi_\nu^\ast \nu_R]$
is acceptable
because the scale of $Y_\nu$ is not necessarily
to be extremely small~\cite{Davidson:2009ha, Kanemura:2013qva}%
\footnote{
 If the $Z_2^\prime$ is broken
not softly but spontaneously~\cite{Ref:nuTHDM-D},
the scale of $Y_\nu$ is constrained
to be extremely small~\cite{Ref:nuTHDM-const}.
}.
 When we introduce $\Phi_2$
in addition to $\Phi$ in the SM,
another softly-broken $Z_2$ symmetry is imposed
such that only $\Phi_2$ couples with $\ell_R$
in order to forbid the flavor changing neutral current%
~\cite{Glashow:1976nt, Barger:1989fj, Ref:THDM-2009}.
 Then,
$\Phi_2$ provides the diagonal Yukawa matrix,
whose diagonal elements $y_\ell$
are proportional to the charged lepton masses $m_\ell^{}$.
 In contrast,
$s_L^+$ gives the antisymmetric Yukawa matrix $Y_A^s$
while $s^0$, $s^{++}$, and $\Delta$ have symmetric Yukawa matrices
$Y_S^0$, $Y_S^s$, and $Y_S^\Delta$, respectively.
 Notice that
$s^0$ and $\Delta^0$ with $\text{L\#} = -2$
must not have the vacuum expectation values
because of the lepton number conservation.
 When $\nu_L^{}$ is connected to $\nu_R^{}$
by using combinations of the charged current interaction
and Yukawa interactions in Table~\ref{tab:particle-I},
these combinations correspond to
some models for generating $m_\text{D}^{}$.
 As long as we concentrate on the flavor structure,
it is not necessary to specify how the scalar lines are closed.
 If we specify that,
it gives a certain model.



\begin{figure}[t]
 \begin{minipage}{0.45\hsize}
  \vspace*{1mm}
  \begin{center}
   \includegraphics[scale=0.7]{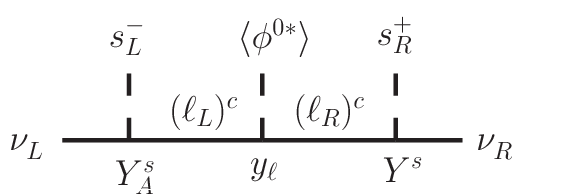}
   \vspace*{-3mm}
   \caption{
    The diagram for the flavor structure in eq.~\eqref{eq:1}.
   }
   \label{fig:NM-type}
  \end{center}
 \end{minipage}
\hspace*{8mm}
 \begin{minipage}{0.45\hsize}
  \begin{center}
   \includegraphics[scale=0.7]{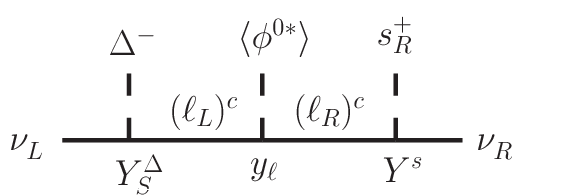}
   \vspace*{-3mm}
   \caption{
    The diagram for the flavor structure in eq.~\eqref{eq:2}.
   }
   \label{fig:Dirac-2}
  \end{center}
 \end{minipage}
\vspace*{5mm}
\end{figure}

\begin{figure}[t]
 \begin{minipage}{0.45\hsize}
  \vspace*{1mm}
  \begin{center}
   \includegraphics[scale=0.7]{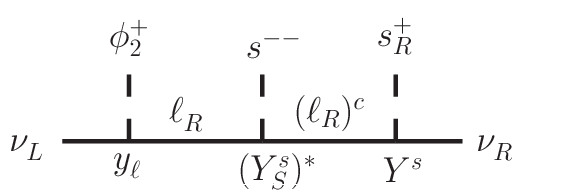}
   \vspace*{-3mm}
   \caption{
    The diagram for the flavor structure in eq.~\eqref{eq:3}.
   }
   \label{fig:Dirac-3}
  \end{center}
 \end{minipage}
\hspace*{8mm}
 \begin{minipage}{0.45\hsize}
  \begin{center}
   \includegraphics[scale=0.7]{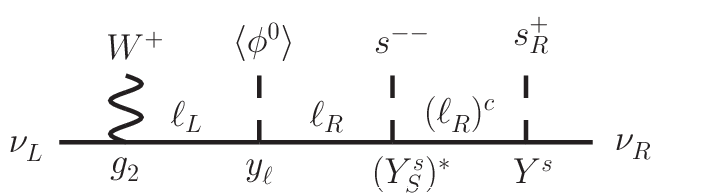}
   \vspace*{-7mm}
   \caption{
    The diagram for the flavor structure in eq.~\eqref{eq:4}.
   }
   \label{fig:Dirac-4}
  \end{center}
 \end{minipage}
\vspace*{5mm}
\end{figure}

\begin{figure}[t]
 \begin{minipage}{0.45\hsize}
  \vspace*{1mm}
  \begin{center}
   \includegraphics[scale=0.7]{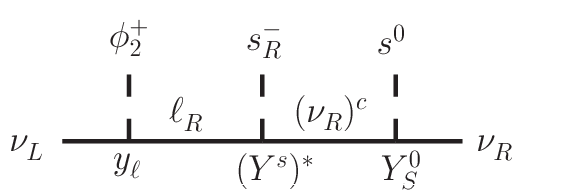}
   \vspace*{-3mm}
   \caption{
    The diagram for the flavor structure in eq.~\eqref{eq:5}.
   }
  \label{fig:Dirac-5}
  \end{center}
 \end{minipage}
\hspace*{8mm}
 \begin{minipage}{0.45\hsize}
  \begin{center}
   \includegraphics[scale=0.7]{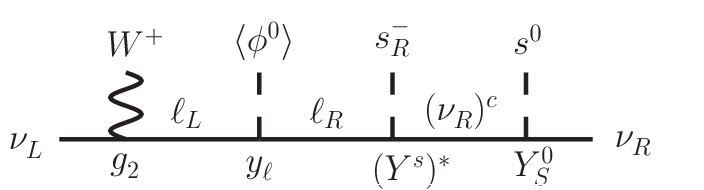}
   \vspace*{-7mm}
   \caption{
    The diagram for the flavor structure in eq.~\eqref{eq:6}.
   }
   \label{fig:Dirac-6}
  \end{center}
 \end{minipage}
\vspace*{5mm}
\end{figure}

\begin{figure}[t]
\begin{center}
\includegraphics[scale=0.7]{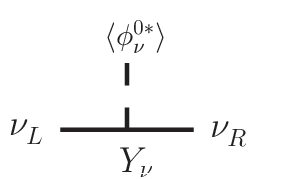}
\caption{
 The diagram for the flavor structure in eq.~\eqref{eq:7}.
}
\label{fig:nuTHDM-type}
\end{center}
\end{figure}

 Each of fermions%
~$( \ell_L, \ell_R, (\ell_L)^c, (\ell_R)^c, (\nu_L^{})^c, (\nu_R^{})^c )$
should not be used twice on a fermion line
from $\nu_L^{}$ to $\nu_R^{}$.
 If a fermion is used twice on a line,
removal of the structure between them gives a simpler line,
which is expected to have a larger contribution to $m_\text{D}^{}$.
 Fermions $(\nu_L^{})^c$ and $(\nu_R^{})^c$
must not appear at the same time on the fermion line
because the structure between them
gives a simpler mechanism to generate $m_\text{D}^{}$.
 Similarly,
when both of $\ell_L$ and $\ell_R$~($(\ell_L)^c$ and $(\ell_R)^c$)
exist on a fermion line,
they should be next to each other.
 If there is a structure between them,
the replacement of the structure with $y_\ell$
provides a simpler mechanism,
whose contribution to $m_\text{D}^{}$
is expected to be larger%
\footnote{
 Since $y_\ell$ includes $y_\tau \sim 10^{-2}$,
the contribution with $y_\ell$ would not be negligible
although $y_e \sim 10^{-6}$ is rather small.
}.
 One might think that
$\ell_L$ should appear next to $\nu_L^{}$
because of the charged current interaction.
 We do not take the restriction
because there is a counter example~(the Zee model~\cite{Zee:1980ai})
for the Majorana neutrino mass.
 However,
we see that
$\ell_L$ always appears next to $\nu_L$
as a result of our analyses for the Dirac neutrino mass.
 Assuming that
the neutrino mass matrix is generated
by a single mechanism%
~(a pattern of alignments of Yukawa matrices),
we find there are seven possibilities
for the flavor structure as follows:
\begin{eqnarray}
m_\text{D}^{}
 &\propto& Y_A^s\, y_\ell^{}\, Y^s, 
\label{eq:1}\\
m_\text{D}^{}
 &\propto& Y_S^\Delta\, y_\ell^{}\, Y^s, 
\label{eq:2}\\
m_\text{D}^{}
 &\propto& y_\ell^{}\, (Y_S^s)^\ast\, Y^s, 
\label{eq:3}\\
m_\text{D}^{}
 &\propto& g_2^{}\, y_\ell^{}\, (Y_S^s)^\ast\, Y^s, 
\label{eq:4}\\
m_\text{D}^{}
 &\propto& y_\ell^{}\, (Y^s)^\ast\, Y_S^0, 
\label{eq:5}\\
m_\text{D}^{}
 &\propto& g_2^{}\, y_\ell^{}\, (Y^s)^\ast\, Y_S^0, 
\label{eq:6}\\
m_\text{D}^{}
 &\propto& Y_\nu ,
\label{eq:7}
\end{eqnarray}
where $g_2$ is the $\SU(2)_L$ gauge coupling constant,
and Yukawa matrices
($Y_A^s$, $y_\ell^{}$, $Y^s$, $Y_S^\Delta$, $Y_S^s$, $Y_S^0$, $Y_\nu$)
are defined in Table~\ref{tab:particle-I}.
 Diagrams of fermion lines for eqs.~\eqref{eq:1}-\eqref{eq:7}
are presented in Figs.~\ref{fig:NM-type}-\ref{fig:nuTHDM-type},
respectively.
 Since the charged current interaction
does not depend on the flavor,
eqs.~\eqref{eq:3} and \eqref{eq:4}
(eqs.~\eqref{eq:5} and \eqref{eq:6})
have the same flavor structure.
 However,
eqs.~\eqref{eq:3} and \eqref{eq:4}
(eqs.~\eqref{eq:5} and \eqref{eq:6})
correspond to different models
because the second Higgs doublet field $\Phi_2$
is required to be introduced
for eq.~\eqref{eq:3}~(eq.~\eqref{eq:5})%
\footnote{
 Although the contribution from eq.~\eqref{eq:4}~(eq.~\eqref{eq:6})
still exists even if $\Phi_2$ is introduced,
it must not be the dominant one
unless the fine tuning of parameters.
 See also Figs.~\ref{fig:loop-1} and \ref{fig:loop-2}
in Appendix~\ref{sec:app-A}.
}.


 The model in Refs.~\cite{1loopDirac, Kanemura:2011jj}
is an example for the structure in Fig.~\ref{fig:NM-type}.
 The scalar lines are connected
via the interaction $\mu^2 [s_L^+ s_R^-]$,
where $\mu$ is the soft-breaking parameter for $Z_2^\prime$.
For Fig.~\ref{fig:nuTHDM-type},
explicit models can be found in Refs.~\cite{Ref:nuTHDM-D, Davidson:2009ha}.
 The $Z_2^\prime$ symmetry
can be softly broken by $\mu^2 [\Phi^\dagger \Phi_\nu]$.
 For the other five structures
in Figs.~\ref{fig:Dirac-2}-\ref{fig:Dirac-6},
explicit models have not been known.
 In Appendix~\ref{sec:app-A},
we show an example to close scalar lines
for each of Figs.~\ref{fig:Dirac-2}-\ref{fig:Dirac-6}.

\begin{table}[t]
\begin{tabular}{c||c|c|c|c|c|c}
Scalar
 & $\SU(2)_L$
 & $\U(1)_Y$
 & \ L\# \
 & $Z_2^\prime$
 & Yukawa
 & Note
\\[0.5mm]
\hline
\hline
$s^0$
 & ${\bf\underline{1}}$
 & $0$
 & $-2$
 & \ Even \
 & $
    (Y_S^0)_{ij}
     \Bigl[
      \overline{(\nu_{iR}^{})^c}\, \nu_{jR}^{}\, s^0
     \Bigr]
   $
 & Symmetric
\\[1mm]
\hline
$s_L^+$
 & ${\bf\underline{1}}$
 & $1$
 & $-2$
 & Even
 & $
    (Y_A^s)_{\ell\ellp}
     \Bigl[
      \overline{L_\ell^{}}\, \epsilon L_{\ell^\prime}^c\, s_L^-
     \Bigr]
   $
 & Antisymmetric
\\[1mm]
\hline
$s_R^+$
 & ${\bf\underline{1}}$
 & $1$
 & $-2$
 & Odd
 & $
    (Y^s)_{\ell i}
     \Bigl[
      \overline{(\ell_R)^c}\, \nu_{iR}^{}\, s_R^+
     \Bigr]
   $
 & Arbitrary
\\[1mm]
\hline
$s^{++}$
 & ${\bf\underline{1}}$
 & $2$
 & $-2$
 & Even
 & $
    (Y_S^s)_{\ell\ellp}
     \Bigl[
      \overline{(\ell_R^{})^c}\, \ell^\prime_R\, s^{++}
     \Bigr]
   $
 & Symmetric
\\[1mm]
\hline
$
\Phi_\nu
=
\begin{pmatrix}
 \phi_\nu^+\\
 \phi_\nu^0
\end{pmatrix}
$
 & ${\bf\underline{2}}$
 & $\displaystyle\frac{1}{\,2\,}$
 & $0$
 & Odd
 & $
    (Y_\nu^{})_{\ell i}
    \Bigl[
     \overline{L_\ell^{}}\, \epsilon\, \Phi_\nu^\ast\, \nu_{iR}^{}
    \Bigr]
   $
 & Arbitrary
\\
\hline
$
\Phi_2
=
\begin{pmatrix}
 \phi_2^+\\
 \phi_2^0
\end{pmatrix}
$
 & ${\bf\underline{2}}$
 & $\displaystyle\frac{1}{\,2\,}$
 & $0$
 & Even
 & $
    y_\ell^{}
    \Bigl[
     \overline{L_\ell^{}}\, \Phi_2\, \ell_R^{}
    \Bigr]
   $
 & Diagonal
\\
\hline
$
\Delta
=
\begin{pmatrix}
 \displaystyle
 \frac{\ \ \Delta^+}{\sqrt{2}}
  & \Delta^{++}\\[3mm]
 \Delta^0
  & \displaystyle -\frac{\ \ \Delta^+}{\sqrt{2}}
\end{pmatrix}
$
 & ${\bf\underline{3}}$
 & $1$
 & $-2$
 & Even
 & $
    (Y_S^\Delta)_{\ell\ell^\prime}
    \Bigl[
     \overline{L_\ell^{}}\, \Delta^\dagger \epsilon\,
     L_{\ell^\prime}^c
    \Bigr]
   $
 & Symmetric
\end{tabular}
\caption
{
 Scalar fields which have Yukawa interactions with leptons.
}
\label{tab:particle-I}
\vspace*{5mm}
\end{table}

 Next,
we classify models
that have the dark matter candidate.
 In addition to $\nu_{iR}^{}$ and
scalar fields in Table~\ref{tab:particle-I},
we introduce $\psi_{iR}^0$
as singlet fermions under the SM gauge group.
 The number of $\psi_R$ is equal to or more than 3
in order to obtain three neutrino masses.
 The lepton number $\text{L\#}=0$ is assigned to $\psi_R^0$
in contrast to $\nu_R$ with $\text{L\#=1}$.
 The Majorana mass term
$(1/2) M_\psi \Bigl[ \overline{(\psi_R^0)^c} \psi_R^0 \Bigr]$
is not forbidden by the lepton number conservation.
 For our classification,
we use Yukawa interactions between $\psi_R^0$ and leptons
by introducing scalar fields
listed in Table~\ref{tab:particle-II}.
 Representations of $s_2^0$, $s_2^+$, and $\eta$
under the SM gauge group
are the same as those of $s^0$, $s_L^+$, and $\Phi$~($\Phi_2$),
respectively.
 The scalar fields in Table~\ref{tab:particle-II}
have $\text{L\#}=-1$
while L\# of $s^0$, $s_L^+$, and $\Phi$~($\Phi_2$)
are even numbers.
 For concreteness,
we take $s_2^0$ as an odd field under $Z_2^\prime$
while $s_2^+$, $\eta$, and $\psi_R^0$
are taken as even fields%
\footnote{
 The opposite assignment is also acceptable.
}.
 Notice that
there appears an unbroken $Z_2$ symmetry,
where $\psi_R^0$ and scalar fields in Table~\ref{tab:particle-II}
are odd due to the L\# assignments%
\footnote{
 The global $\U(1)_\text{F\#+L\#}$ symmetry,
where F\# denotes the fermion number,
is broken down into the $Z_2$ symmetry
by the Majorana mass term of $\psi_R^0$.
 Each field has the $Z_2$ parity $(-1)^\text{F\#+L\#}$.
 At the same time,
the L\# conservation protects the $Z_2$ breaking
because $Z_2$-odd scalar fields have non-zero L\#.
}.
 Since the lightest $Z_2$-odd particle is stable,
it can be the dark matter candidate~(if it is electrically neutral).

\begin{figure}[t]
 \begin{minipage}{0.5\hsize}
  \vspace*{1mm}
  \begin{center}
   \includegraphics[scale=0.5]{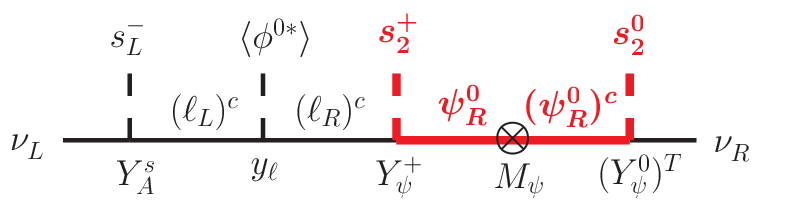}
   \vspace*{-3mm}
   \caption{
    The diagram for the flavor structure in eq.~\eqref{eq:Z2-1}.
    Bold red lines are for odd particles
    of the unbroken $Z_2$ symmetry.
   }
   \label{fig:Z2-1}
  \end{center}
 \end{minipage}
\end{figure}

\begin{figure}[t]
 \begin{minipage}{0.4\hsize}
  \vspace*{7mm}
  \begin{center}
   \includegraphics[scale=0.5]{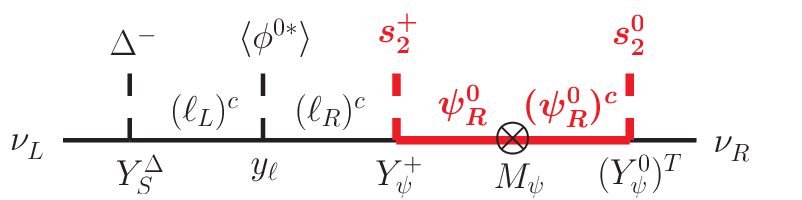}
   \vspace*{-10mm}
   \caption{
    The diagram for the flavor structure in eq.~\eqref{eq:Z2-2}.
    Bold red lines are for odd particles
    of the unbroken $Z_2$ symmetry.
   }
   \label{fig:Z2-2}
  \end{center}
 \end{minipage}
\hspace*{3mm}
 \begin{minipage}{0.45\hsize}
  \begin{center}
   \includegraphics[scale=0.5]{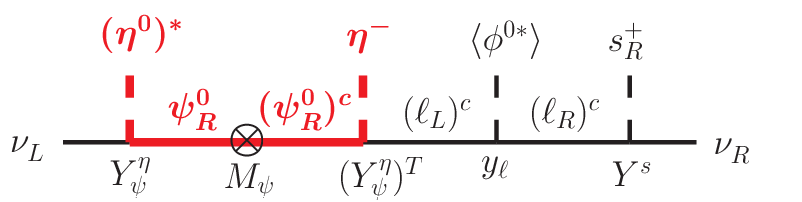}
   \vspace*{-3mm}
   \caption{
    The diagram for the flavor structure in eq.~\eqref{eq:Z2-3}.
    Bold red lines are for odd particles
    of the unbroken $Z_2$ symmetry.
   }
   \label{fig:Z2-3}
  \end{center}
 \end{minipage}
\vspace*{5mm}
\end{figure}

\begin{figure}[t]
 \begin{minipage}{0.4\hsize}
  \vspace*{9mm}
  \begin{center}
   \includegraphics[scale=0.5]{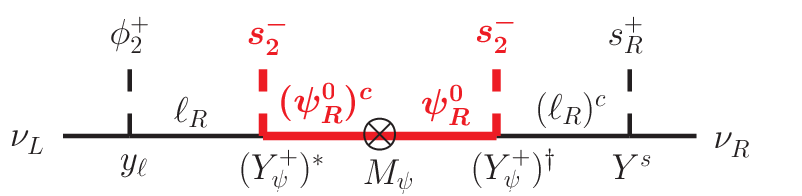}
   \vspace*{-11mm}
   \caption{
    The diagram for the flavor structure in eq.~\eqref{eq:Z2-4}.
    Bold red lines are for odd particles
    of the unbroken $Z_2$ symmetry.
   }
   \label{fig:Z2-4}
  \end{center}
 \end{minipage}
\hspace*{3mm}
 \begin{minipage}{0.45\hsize}
  \vspace*{1mm}
  \begin{center}
   \includegraphics[scale=0.5]{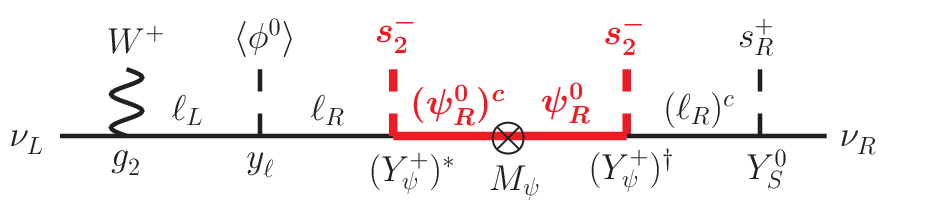}
   \vspace*{-11mm}
   \caption{
    The diagram for the flavor structure in eq.~\eqref{eq:Z2-5}.
    Bold red lines are for odd particles
    of the unbroken $Z_2$ symmetry.
   }
   \label{fig:Z2-5}
  \end{center}
 \end{minipage}
\vspace*{5mm}
\end{figure}

\begin{figure}[t]
 \begin{minipage}{0.4\hsize}
  \vspace*{9mm}
  \begin{center}
   \includegraphics[scale=0.5]{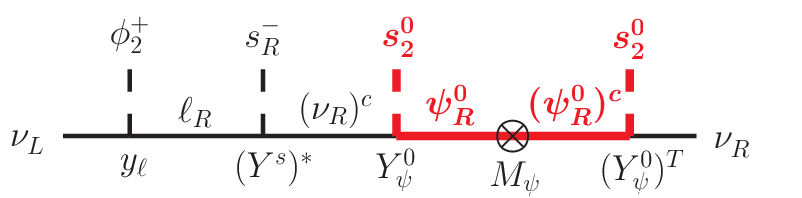}
   \vspace*{-10mm}
   \caption{
    The diagram for the flavor structure in eq.~\eqref{eq:Z2-6}.
    Bold red lines are for odd particles
    of the unbroken $Z_2$ symmetry.
   }
   \label{fig:Z2-6}
  \end{center}
 \end{minipage}
\hspace*{3mm}
 \begin{minipage}{0.45\hsize}
  \vspace*{1mm}
  \begin{center}
   \includegraphics[scale=0.5]{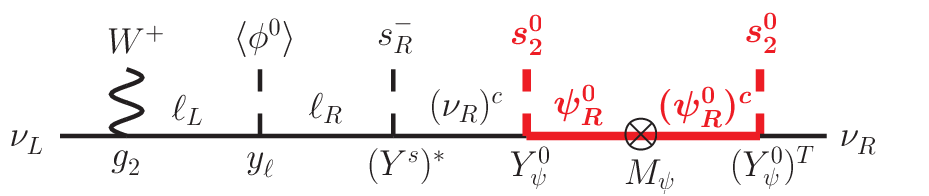}
   \vspace*{-10mm}
   \caption{
    The diagram for the flavor structure in eq.~\eqref{eq:Z2-7}.
    Bold red lines are for odd particles
    of the unbroken $Z_2$ symmetry.
   }
   \label{fig:Z2-7}
  \end{center}
 \end{minipage}
\vspace*{5mm}
\end{figure}

\begin{figure}[t]
 \begin{minipage}{0.4\hsize}
  \vspace*{8mm}
  \begin{center}
   \includegraphics[scale=0.5]{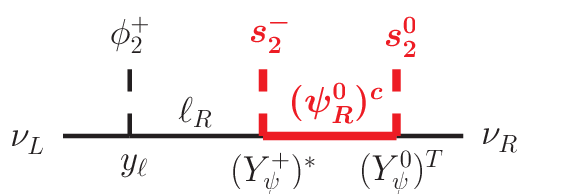}
   \vspace*{-4mm}
   \caption{
    The diagram for the flavor structure in eq.~\eqref{eq:Z2-8}.
    Bold red lines are for odd particles
    of the unbroken $Z_2$ symmetry.
   }
   \label{fig:Z2-8}
  \end{center}
 \end{minipage}
\hspace*{3mm}
 \begin{minipage}{0.45\hsize}
  \begin{center}
   \includegraphics[scale=0.5]{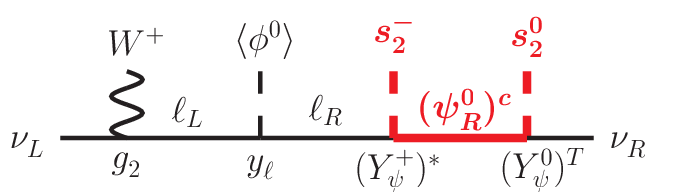}
   \vspace*{-3mm}
   \caption{
    The diagram for the flavor structure in eq.~\eqref{eq:Z2-9}.
    Bold red lines are for odd particles
    of the unbroken $Z_2$ symmetry.
   }
   \label{fig:Z2-9}
  \end{center}
 \end{minipage}
\vspace*{5mm}
\end{figure}

\begin{figure}[t]
 \begin{minipage}{0.4\hsize}
  \vspace*{7mm}
  \begin{center}
   \includegraphics[scale=0.5]{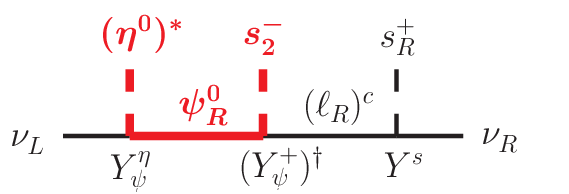}
   \vspace*{-3mm}
   \caption{
    The diagram for the flavor structure in eq.~\eqref{eq:Z2-10}.
    Bold red lines are for odd particles
    of the unbroken $Z_2$ symmetry.
   }
   \label{fig:Z2-10}
  \end{center}
 \end{minipage}
\hspace*{3mm}
 \begin{minipage}{0.45\hsize}
  \begin{center}
   \includegraphics[scale=0.5]{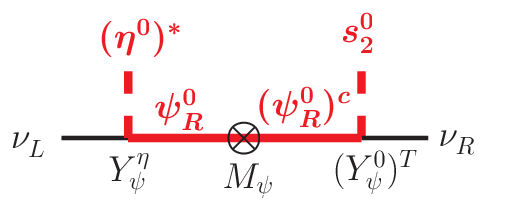}
   \vspace*{-3mm}
   \caption{
    The diagram for the flavor structure in eq.~\eqref{eq:Z2-11}.
    Bold red lines are for odd particles
    of the unbroken $Z_2$ symmetry.
   }
   \label{fig:GS-type}
  \end{center}
 \end{minipage}
\vspace*{5mm}
\end{figure}

 Let us consider fermion lines
to connect $\nu_L$ with $\nu_R$
by using also the $Z_2$-odd particles.
 Similarly to the case without the $Z_2$-odd particles,
$\psi_R^0$ and $(\psi_R^0)^c$ should not appear twice
on a fermion line.
 When both of them appear,
they should be next to each other
because of their mass term.
 In addition to eqs.~\eqref{eq:1}-\eqref{eq:7},
we obtain the following eleven combinations:
\begin{eqnarray}
m_\text{D}^{}
&\propto&
 Y_A^s\, y_\ell^{}\, Y_\psi^+\, M_\psi^{-1}\, (Y_\psi^0)^T ,
\label{eq:Z2-1}
\\
%
%
m_\text{D}^{}
&\propto&
 Y_S^\Delta\, y_\ell^{}\, Y_\psi^+\, M_\psi^{-1}\, (Y_\psi^0)^T ,
\label{eq:Z2-2}
\\
%
%
m_\text{D}^{}
&\propto&
 Y_\psi^\eta\, M_\psi^{-1}\, (Y_\psi^\eta)^T\, y_\ell^{}\, Y^s ,
\label{eq:Z2-3}
\\
%
%
m_\text{D}^{}
&\propto&
 y_\ell^{}\, (Y_\psi^+)^\ast\, M_\psi^{-1}\, (Y_\psi^+)^\dagger\, Y^s ,
\label{eq:Z2-4}
\\
%
%
m_\text{D}^{}
&\propto&
 g_2^{}\, y_\ell^{}\,
 (Y_\psi^+)^\ast\, M_\psi^{-1}\, (Y_\psi^+)^\dagger\, Y^s ,
\label{eq:Z2-5}
\\
%
%
m_\text{D}^{}
&\propto&
 y_\ell^{}\, (Y^s)^\ast\, Y_\psi^0\, M_\psi^{-1}\, (Y_\psi^0)^T ,
\label{eq:Z2-6}
\\
%
%
m_\text{D}^{}
&\propto&
 g_2^{}\, y_\ell^{}\, (Y^s)^\ast\, Y_\psi^0\, M_\psi^{-1}\, (Y_\psi^0)^T ,
\label{eq:Z2-7}
\\
%
%
m_\text{D}^{}
&\propto&
 y_\ell^{}\, (Y_\psi^+)^\ast\, (Y_\psi^0)^T ,
\label{eq:Z2-8}
\\
%
%
m_\text{D}^{}
&\propto&
 g_2^{}\, y_\ell^{}\, (Y_\psi^+)^\ast\, (Y_\psi^0)^T ,
\label{eq:Z2-9}
\\
%
%
m_\text{D}^{}
&\propto&
 Y_\psi^\eta\, (Y_\psi^+)^\dagger\, Y^s ,
\label{eq:Z2-10}
\\
%
%
m_\text{D}^{}
&\propto&
 Y_\psi^\eta\, M_\psi^{-1}\, (Y_\psi^0)^T ,
\label{eq:Z2-11}
\end{eqnarray}
where Yukawa matrices
$Y_\psi^0$, $Y_\psi^+$, and $Y_\psi^\eta$
are defined in Table~\ref{tab:particle-II}.
 Fermion lines for eqs.~\eqref{eq:Z2-1}-~\eqref{eq:Z2-11}
are shown in Figs.~\ref{fig:Z2-1}-\ref{fig:GS-type}.
 The flavor structures of
eqs.~\eqref{eq:Z2-4}, \eqref{eq:Z2-6}, and \eqref{eq:Z2-8}
are the same as those of
eqs.~\eqref{eq:Z2-5}, \eqref{eq:Z2-7}, and \eqref{eq:Z2-9},
respectively.
 They correspond to different models
because eqs.~\eqref{eq:Z2-4}, \eqref{eq:Z2-6}, and \eqref{eq:Z2-8}
require $\Phi_2$.

 Scalar lines in Fig.~\ref{fig:GS-type} can be connected
via $\mu[\Phi^\dagger \eta (s_2^0)^\ast]$
as we see in Ref.~\cite{Gu:2007ug}%
~(See also Ref.~\cite{Farzan:2012sa}).
 For the other ten structures in Figs.~\ref{fig:Z2-1}-\ref{fig:Z2-10},
explicit models have not been known.
 An example to close scalar lines
for each of Figs.~\ref{fig:Z2-1}-\ref{fig:Z2-10}
is presented in Appendix~\ref{sec:app-B}.


 As a result,
structures in eqs.~\eqref{eq:1}-\eqref{eq:7}
and eqs.~\eqref{eq:Z2-1}-~\eqref{eq:Z2-11}
can be classified into seven groups as follows:
\begin{eqnarray}
\text{Group-I}
:&&\!\!\!
m_\text{D}^{}
 \propto Y_A^s\, y_\ell^{}\, X^s ,
 \qquad\qquad
 X^s = Y^s, \ Y_\psi^+\, M_\psi^{-1}\, (Y_\psi^0)^T ,
\label{eq:G1}
\\
%
%
\text{Group-II}
:&&\!\!\!
m_\text{D}^{}
 \propto X_{SL}\, y_\ell^{}\, X^s ,
\nonumber\\
&&\!\!\!
 \{ X_{SL} , \ X^s \}
 = \{ Y_S^\Delta , \ Y^s \} , \
   \{ Y_\psi^\eta\, M_\psi^{-1} (Y_\psi^\eta)^T , \ Y^s \} , \
   \{ Y_S^\Delta , \ Y_\psi^+\, M_\psi^{-1}\, (Y_\psi^0)^T \} ,
\hspace*{5mm}
\label{eq:G2}
\\
%
%
\text{Group-III}
:&&\!\!\!
m_\text{D}^{}
 \propto y_\ell^{}\, X_{SR}^\ast\, Y^s ,
 \qquad\quad
 X_{SR} = Y_S^s , \ (Y_\psi^+)^\ast\, M_\psi^{-1}\, (Y_\psi^+)^\dagger ,
\label{eq:G3}
\\
%
%
\text{Group-IV}
:&&\!\!\!
m_\text{D}^{}
 \propto y_\ell^{}\, (Y^s)^\ast\, X_{S\nu} ,
 \qquad X_{S\nu} = Y_S^0 , \ Y_\psi^0\, M_\psi^{-1}\, (Y_\psi^0)^T ,
\label{eq:G4}
\\
%
%
\text{Group-V}
:&&\!\!\!
m_\text{D}^{}
 \propto y_\ell^{}\, X_\psi ,
 \qquad\qquad\quad X_\psi = (Y_\psi^+)^\ast\, (Y_\psi^0)^T ,
\label{eq:G5}
\\
%
%
\text{Group-VI}
:&&\!\!\!
m_\text{D}^{}
 \propto X_\psi^\eta\, Y^s ,
 \qquad\qquad \ \, X_\psi^\eta = Y_\psi^\eta\, (Y_\psi^+)^\dagger ,
\label{eq:G6}
\\
%
%
\text{Group-VII}
:&&\!\!\!
m_\text{D}^{}
 \propto X_\nu ,
 \qquad\qquad\qquad
 X_\nu = Y_\nu , \ (Y_\psi^\eta)\, M_\psi^{-1}\, (Y_\psi^0)^T .
\label{eq:G7}
\end{eqnarray}
 Notice that
$X_{SL}$, $X_{SR}$, and $X_{S\nu}$ are symmetric matrices.
 Structures of these groups
are given in terms of interactions between leptons%
~(new fermions are hidden in interactions $X$)
and cannot be simpler.
 Therefore,
they cannot be included in any other groups,
and they correspond to independent models.
 Models in Refs.~\cite{1loopDirac, Kanemura:2011jj, Chen:2012baa}
are included in the \text{Group-I}.
 The Group-VII contains
models in Refs.~\cite{Gu:2007ug, Farzan:2012sa, Okada:2014vla}.
 Although the flavor structure
in the Dirac seesaw mechanism~\cite{DSeesaw}
is the same as the structure of the Group-VII,
we do not put it into the group.
 This is because the Dirac seesaw mechanism
has no charged scalar,
which contributes to charged lepton decays,
unlike models in Refs.~\cite{Gu:2007ug, Farzan:2012sa, Okada:2014vla}.
 Since models in Ref.~\cite{1loopDirac-LR}
is given by extending the gauge group of the SM,
they are not included in the above seven groups.

\begin{table}[t]
\renewcommand{\arraystretch}{1.5}
\begin{tabular}{c||c|c|c|c|c|c}
Scalar
 & $\SU(2)_L$
 & $\U(1)_Y$
 & L\#
 & $Z_2^\prime$
 & Yukawa
 & Note
\\[0.5mm]
\hline
\hline
$\red{s_2^0}$
 & ${\bf\underline{1}}$
 & $0$
 & $-1$
 & Odd
 & $
    (Y_\psi^0)_{ij}
     \Bigl[
      \overline{(\nu_{iR}^{})^c}\, \red{\psi_{jR}^0}\, \red{s_2^0}
     \Bigr]
   $
 & Arbitrary
\\[1mm]
\hline
$\red{s_2^+}$
 & ${\bf\underline{1}}$
 & $1$
 & $-1$
 & Even
 & $
    (Y_\psi^+)_{\ell i}
     \Bigl[
      \overline{(\ell_R^{})^c}\, \red{\psi_{iR}^0}\, \red{s_2^+}
     \Bigr]
   $
 & Arbitrary
\\[1mm]
\hline
\renewcommand{\arraystretch}{1}
$
\red{\eta}
=
\begin{pmatrix}
 \red{\eta^+}\\
 \red{\eta^0}
\end{pmatrix}
$
 & ${\bf\underline{2}}$
 & $\displaystyle\frac{1}{\,2\,}$
 & $-1$
 & Even
 & $
    (Y_\psi^\eta)_{\ell i}
    \Bigl[
     \overline{L_\ell^{}}\, \epsilon\, \red{\eta^\ast}\, \red{\psi_{iR}^0}
    \Bigr]
   $
 & Arbitrary
\end{tabular}
\caption
{
 Scalar fields which have Yukawa interactions
with $\psi_R^0$ and leptons.
}
\label{tab:particle-II}
\vspace*{5mm}
\end{table}

\section{Discussion}


 Let us discuss
how we can test these groups
in eqs.~\eqref{eq:G1}-\eqref{eq:G7}.
 The simplest test is the search for $0\nu\beta\beta$,
where the conservation of L\# is violated by two units.
 If the decay is observed,
all groups in eqs.~\eqref{eq:G1}-\eqref{eq:G7} will be excluded
because they are given by assuming the L\# conservation.



 By taking the basis where $\nu_{iR}^{}$ are mass-eigenstates,
the Dirac neutrino mass matrix $m_\text{D}^{}$
can be expressed as
$m_\text{D}^{} = U_\MNS\, \text{diag}(m_1, m_2, m_3)$,
where $m_i$~($i=1\text{-}3$) are neutrino mass eigenvalues.
 The case of $m_1 < m_3$ is referred to as
the normal mass ordering~(NO)
while $m_3 < m_1$ is called as the inverted mass ordering~(IO).
 The mixing matrix $U_\MNS$
is the so-called Maki-Nakagawa-Sakata~(MNS) matrix~\cite{Maki:1962mu},
which can be parameterized as
\begin{eqnarray}
U_\MNS
=
 \begin{pmatrix}
  1 & 0 & 0\\
  0 & c_{23} & s_{23} \\
  0 & -s_{23} & c_{23}
 \end{pmatrix}
 \begin{pmatrix}
  c_{13} & 0 & s_{13} e^{-i\delta} \\
  0 & 1 & 0\\
  -s_{13} e^{i\delta} & 0 & c_{13}
 \end{pmatrix}
 \begin{pmatrix}
  c_{12} & s_{12} & 0\\
  -s_{12} & c_{12} & 0\\
  0 & 0 & 1
 \end{pmatrix} ,
\end{eqnarray}
where $c_{ij} \equiv \cos\theta_{ij}$
and $s_{ij} \equiv \sin\theta_{ij}$.
 For \text{Group-I}~(
$m_\text{D}^{} \propto Y_A^s\, y_\ell^{}\, X^s$),
we see that $\Det(m_\text{D}^{}) \propto \Det(Y_A) = 0$.
 Then,
the smallest eigenvalue must be zero,
namely $m_1 = 0$ or $m_3 = 0$.
 The direct measurement of the absolute neutrino mass
can be achieved at the KATRIN experiment~\cite{Osipowicz:2001sq},
whose expected sensitivity is
$0.35\,\eV$ at $5\sigma$ confidence level.
 The Group-I is excluded
if the experiment gives an affirmative result.
 Cosmological observations
put the indirect bound
$\sum_i m_i < 0.23\,\eV$~($90\,\%$ confidence level)~\cite{Ade:2015xua},
and the future experiments
are expected to have the sensitivity to
$\sum_i m_i = {\mathcal O}(0.01)\,\eV$~\cite{Abazajian:2013oma}.
 If $\sum_i m_i \lesssim 0.1\,\eV$ is excluded,
we see that the lightest neutrino mass is not zero,
and consequently the Group-I is excluded.
 We have the same conclusion
if exclusion of $\sum_i m_i \lesssim 0.06\,\eV$ is achieved
in addition to determination of IO
in neutrino oscillation experiments~\cite{Blennow:2013oma}.



 The matrix $X_\psi$ for the Group-V%
~($m_\text{D}^{} \propto y_\ell\, X_\psi$)
gives the four-fermion interaction
\begin{eqnarray}
{\mathcal L}_\text{4-fermi}
=
 \left( \frac{1}{16\pi^2} \right)^n
 \frac{1}{\Lambda^2}
 (X_\psi)_{\ell i} (X_\psi^\dagger)_{j\ellp}
 \Bigl[ \overline{\ell_R} \gamma_\mu \nu_{iR}^{} \Bigr]
 \Bigl[ \overline{\nu_{jR}^{}} \gamma^\mu \ellp_R \Bigr],
\end{eqnarray}
where $\Lambda$ is the energy scale of the new physics.
 If we use $X_{\psi} = (Y_\psi^+)^\ast (Y_\psi^0)^T$ as an example,
the four-fermion interaction is obtained at the one-loop level~($n=1$).
 The interaction causes
$\ell \to \ellp_R \nu_{iR}^{} \overline{\nu_{jR}^{}}$,
which affect to $\ell \to \ellp \nu \overline{\nu}$
in addition to
$\ell \to \ellp_L \nu_{\ell L}^{} \overline{\nu_{\ellp L}^{}}$
via the charged current interaction.
 Since we do not measure neutrino species,
contributions from $X_\psi$ are summed up as
$(X_\psi X_\psi^\dagger)_{\ell\ell} (X_\psi X_\psi^\dagger)_{\ellp\ellp}$.
 The Fermi coupling constant $G_F$
is given by measuring $\mu \to e \nu\overline{\nu}$.
 We have
$G_F = G^W \equiv g_2^2/(4\sqrt{2}\,m_W^2)$ in the standard model,
where $g_2$ denotes the $\SU(2)_L$ gauge coupling constant,
and $m_W$ is the $W$ boson mass.
 Although the coupling constants $G_{\tau\ellp}$~($\ellp = e, \mu$)
given by measuring $\tau \to \ellp \nu\overline{\nu}$ in the standard model
is equal to $G_F$,
the deviation from it can exist for the Group-V as
\begin{eqnarray}
G_{\tau\ellp}^2
= G_F^2 + (G_{\tau\ellp}^X)^2 - (G_{\mu e}^X)^2 ,
\quad
%
%
(G_{\ell\ellp}^X)^2
\equiv
 \left( \frac{1}{16\pi^2} \right)^{2n}
 \frac{
       (m_\text{D}^{} m_\text{D}^\dagger)_{\ell\ell}
       (m_\text{D}^{} m_\text{D}^\dagger)_{\ellp\ellp}
      }
      { 8\, \Lambda^4 C_\text{loop}^4 m_\ell^2 m_{\ellp}^2 } ,
\end{eqnarray}
where
$(m_\text{D}^{})_{\ell i} = C_\text{loop} m_\ell (X_\psi)_{\ell i}$.
 Coefficients
$(m_\text{D}^{} m_\text{D}^\dagger)_{\ell\ell}$
are given by
\begin{eqnarray}
(m_\text{D}^{} m_\text{D}^\dagger)_{ee}
&=&
 m_1^2
 + c_{13}^2 s_{12}^2 \Delta m^2_{21}
 + s_{13}^2 \Delta m^2_{31}\\
&=&
 m_1^2 + 7.7\times 10^{-5}\,\eV^2 ,
\\
%
%
(m_\text{D}^{} m_\text{D}^\dagger)_{\mu\mu}
&=&
 m_1^2
 + (
    c_{23}^2 c_{12}^2
    + s_{23}^2 s_{13}^2 s_{12}^2
    - 2 c_{23} s_{23} s_{13} c_{12} s_{12} \cos\delta
   ) \Delta m^2_{21}
 + s_{23}^2 c_{13}^2 \Delta m^2_{31}\\
&=&
 m_1^2 +
 (
  1.3\times 10^{-3}
  - 5.0\times 10^{-6} \cos\delta
 )\,\eV^2 ,
\\
%
%
(m_\text{D}^{} m_\text{D}^\dagger)_{\tau\tau}
&=&
 m_1^2
 + (
    s_{23}^2 c_{12}^2
    + c_{23}^2 s_{13}^2 s_{12}^2
    + 2 c_{23} s_{23} s_{13} c_{12} s_{12} \cos\delta
   ) \Delta m^2_{21}
 + c_{23}^2 c_{13}^2 \Delta m^2_{31}\\
&=&
 m_1^2 +
 (
  1.3\times 10^{-3}
  + 5.0\times 10^{-6} \cos\delta
 )\,\eV^2
\end{eqnarray}
for NO
and
\begin{eqnarray}
(m_\text{D}^{} m_\text{D}^\dagger)_{ee}
&=&
 m_3^2 + \Delta m^2_{13}
 + c_{13}^2 s_{12}^2 \Delta m^2_{21}
 - s_{13}^2 \Delta m^2_{13}\\
&=&
 m_3^2 + 2.4\times 10^{-3}\,\eV^2 ,
\\
%
%
(m_\text{D}^{} m_\text{D}^\dagger)_{\mu\mu}
&=&
 m_3^2 + \Delta m^2_{13}
\nonumber\\
&&\hspace*{0mm}{}
 + (
    c_{23}^2 c_{12}^2
    + s_{23}^2 s_{13}^2 s_{12}^2
    - 2 c_{23} s_{23} s_{13} c_{12} s_{12} \cos\delta
   ) \Delta m^2_{21}
 - s_{23}^2 c_{13}^2 \Delta m^2_{13}\\
&=&
 m_3^2
 +
 (
  1.2\times 10^{-3}
  - 5.0\times 10^{-6} \cos\delta
 )\,\eV^2 ,
\\
%
%
(m_\text{D}^{} m_\text{D}^\dagger)_{\tau\tau}
&=&
 m_3^2 + \Delta m^2_{13}
\nonumber\\
&&\hspace*{0mm}{}
 + (
    s_{23}^2 c_{12}^2
    + c_{23}^2 s_{13}^2 s_{12}^2
    + 2 c_{23} s_{23} s_{13} c_{12} s_{12} \cos\delta
   ) \Delta m^2_{21}
 - c_{23}^2 c_{13}^2 \Delta m^2_{13}\\
&=&
 m_3^2 +
 (
  1.3\times 10^{-3}
  + 5.0\times 10^{-6} \cos\delta
 )\,\eV^2
\end{eqnarray}
for IO\@.
 We used the following values:
\begin{eqnarray}
&&
|\Delta m^2_{32}| = 2.51\times 10^{-3}\,\eV^2%
~\text{\cite{Abe:2015awa}} ,
\quad
\Delta m^2_{21} = 7.46\times 10^{-5}\,\eV^2%
~\text{\cite{Aharmim:2011vm}} ,
\\
&&
\sin^2\theta_{23} = 0.514%
~\text{\cite{Abe:2015awa}} ,
\quad
\sin^2(2\theta_{13}) = 0.084%
~\text{\cite{An:2015rpe}} ,
\quad
\tan^2\theta_{12} = 0.427%
~\text{\cite{Aharmim:2011vm}} ,
\end{eqnarray}
where $\Delta m^2_{ij} \equiv m_i^2 - m_j^2$.
 We see $(G_{\mu e}^X)^2 \gg (G_{\tau\ellp}^X)^2$
due to $1/(m_\ell^2 m_{\ellp}^2)$,
and the Group-V predicts
$G_{\tau e}^2 \simeq G_{\tau\mu}^2 \lesssim G_F^2$.

 Similarly to the Group-V,
the Group-VII~($m_\text{D}^{} \propto X_\nu$)
causes
$\ell \to \ellp_L \nu_{iR}^{} \overline{\nu_{jR}^{}}$
via
\begin{eqnarray}
(G_{\ell\ellp}^X)^2
\equiv
 \left( \frac{1}{16\pi^2} \right)^{2n}
 \frac
 {
  (m_\text{D}^{} m_\text{D}^\dagger)_{\ell\ell}
  (m_\text{D}^{} m_\text{D}^\dagger)_{\ellp\ellp}
 }
 { 8\, \Lambda^4 v^4 (C_\text{loop}^\prime)^4 } ,
\end{eqnarray}
where
$(m_\text{D}^{})_{\ell i}
 = C_\text{loop}^\prime (v/\sqrt{2}) (X_\nu)_{\ell i}$.
 If we take $X_\nu = Y_\nu$ as an example,
the four-fermion interaction is generated at the tree level~($n=0$).
 This contribution is known
for models in Refs.~\cite{Ref:nuTHDM-D, Davidson:2009ha},
which belong to the Group-VII\@.
 We see
$(m_\text{D}^{} m_\text{D}^\dagger)_{ee}
\lesssim (m_\text{D}^{} m_\text{D}^\dagger)_{\mu\mu}
\simeq (m_\text{D}^{} m_\text{D}^\dagger)_{\tau\tau}$
for NO
and
$(m_\text{D}^{} m_\text{D}^\dagger)_{ee}
\gtrsim (m_\text{D}^{} m_\text{D}^\dagger)_{\mu\mu}
\simeq (m_\text{D} m_\text{D}^\dagger)_{\tau\tau}$
for IO\@.
 Therefore,
the Group-VII predicts
$G_{\tau \mu}^2 \gtrsim G_{\tau e}^2 \simeq G_F^2$ for NO
and $G_{\tau \mu}^2 \lesssim G_{\tau e}^2 \simeq G_F^2$ for IO\@.

 Predictions of $G_{\ell\ellp}^2$ for the Group-V and the Group-VII
are summarized in Table~\ref{tab:ltolnn}.
 We do not have predictions for the other five groups
though charged scalars in these groups
can also contribute to $\ell \to \ellp \nu \overline{\nu}$.
 Experimental bounds are shown in Ref.~\cite{Agashe:2014kda} as
\begin{eqnarray}
\frac{ G_{\tau e}^2 }{ G_F^2 }
&=& 1.0029 \pm 0.0046 ,
\\
%
%
\frac{ G_{\tau \mu}^2 }{ G_F^2 }
&=& 0.981 \pm 0.018 .
\end{eqnarray}
 The Babar collaboration~\cite{Aubert:2009qj} gives
\begin{eqnarray}
\frac{ G_{\tau\mu}^2 }{ G_{\tau e}^2 }
= 1.0036 \pm 0.0020 ,
\end{eqnarray}
which results in the world average
$G_{\tau\mu}^2/G_{\tau e}^2  = 1.0018 \pm 0.0014$.
 Since experimental results up to now are
consistent with the prediction in the standard model,
more precise data%
~(at the Belle experiment or the Belle-II experiment~\cite{Abe:2010gxa})
would be desired to test the Group-V and the Group-VII\@.
 If a deviation of $G_{\tau\mu}^2/G_{\tau e}^2$ from unity
is discovered as predicted for the Group-VII,
the group would be tested further
by the determination of the ordering
of neutrino masses~(NO or IO)
in neutrino oscillation experiments~\cite{Blennow:2013oma}.



 For tests of the remaining four groups,
we need discovery of some new scalar particle
at collider experiments%
\footnote{
 In general,
doublet scalar fields affect
the electroweak precision tests.
 However,
their contributions are negligible
if we take degenerate masses
of the charged and the CP-odd Higgs bosons
similarly to the case
in the two Higgs doublet models~(see e.g.\ Ref.~\cite{Kanemura:2011sj}).
 Since singlet and triplet scalar fields in our analyses
do not have vacuum expectation values,
they do not have large contributions
to the electroweak precision tests.
}.
 In the case of discovery of
the doubly charged scalar
that decays into a pair of the same-sign charged leptons,
the Group-II~(see Fig.~\ref{fig:Dirac-2})
and the \text{Group-III}~(see Figs.~\ref{fig:Dirac-3} and \ref{fig:Dirac-4})
would be supported.
 If experiments discover
the charged scalar that dominantly decays into $\tau$,
the particle could be identified as $\phi_2^-$.
 Then,
the Group-III~(see Figs.~\ref{fig:Dirac-3} and \ref{fig:Z2-4})
and the Group-IV~(see Figs.~\ref{fig:Dirac-5} and \ref{fig:Z2-6})
as well as the Group-V~(see Fig.~\ref{fig:Z2-8})
would be preferred.
 The Group-II~(see Fig.~\ref{fig:Z2-3})
and the Group-VI~(see Fig.~\ref{fig:Z2-10})
would be supported
together with the Group-VII~(see Fig.~\ref{fig:GS-type})
if some scalar that comes from
$\eta$~(odd under the unbroken $Z_2$) is discovered.
 Even for the Group-I and the Group-VII,
which can be tested without discovery of new particles,
measurements of decay patterns of the charged scalar
can be utilized for the test
because explicit models for these groups
have predictions for the decay patterns%
~\cite{Kanemura:2011jj, Davidson:2009ha}.


\begin{table}[t]
\begin{tabular}{c||c|c}
{}
 & Group-V 
 & Group-VII
\\\hline
 $\ell \to \ell^\prime \nu \overline{\nu}$
 & $G_{\tau \mu}^2 \simeq G_{\tau e}^2 \lesssim G_F^2$
 & $G_{\tau \mu}^2 \gtrsim G_{\tau e}^2 \simeq G_F^2$ ($m_1 < m_3$)
\\
 {}
 & {}
 & $G_{\tau \mu}^2 \lesssim G_{\tau e}^2 \simeq G_F^2$ ($m_1 > m_3$)
\end{tabular}
\caption
{
 Predictions for deviations from the lepton universality
in cases of the Group-V and the Group-VII\@.
}
\label{tab:ltolnn}
\vspace*{5mm}
\end{table}

\section{Conclusion}

 In this letter,
we have classified new physics models for the Dirac neutrino mass
according to combinations of Yukawa interactions.
 Detail of models is not required for our classification
because we concentrate on the flavor structure
of the neutrino mass matrix,
which is determined only by Yukawa matrices.
 If all possible Yukawa interactions between leptons
are taken into account for our classification,
we have found that there are
seven combinations of them
for the flavor structure of $m_\text{D}^{}$.
 Additional eleven combination of Yukawa interactions appear
if we add singlet-fermions $\psi_{iR}^0$ with $\text{L\#} = 0$
and scalar fields for Yukawa interactions
between $\psi_{iR}^0$ and leptons
in order to obtain the dark matter candidate.
 The dark matter candidate is stabilized
by the unbroken $Z_2$ symmetry,
which appears due to assignments of L\#.
 We have shown that
these combinations can be classified into seven groups.

 If the neutrinoless double beta decay is observed,
these groups are excluded
because the conservation of L\#is assumed.
 The Group-I~($m_\text{D}^{} \propto Y_A^s\, y_\ell^{}\, X^s$)
in eq.~\eqref{eq:G1},
where $Y_A^s$ is an antisymmetric Yukawa matrix,
predicts $\text{min}(m_1, m_3) = 0$.
 Thus,
the Group-I can be tested
by direct~\cite{Osipowicz:2001sq}
and indirect~\cite{Abazajian:2013oma} measurements
of the absolute neutrino mass.
 The Group-V%
~($m_\text{D}^{} \propto y_\ell^{}\, X_\psi$) in eq.~\eqref{eq:G5},
where $y_\ell$ is the diagonal Yukawa matrix for charged lepton masses,
predicts $G_{\tau \mu}^2 \simeq G_{\tau e}^2 \lesssim G_F^2$
for possible deviations
from the lepton universality
in $\ell \to \ellp \nu \overline{\nu}$
due to the interaction with the matrix $X_\psi$.
 The Group-VII%
~($m_\text{D}^{} \propto X_\nu$) in eq.~\eqref{eq:G7}
predicts
$G_{\tau \mu}^2 \gtrsim G_{\tau e}^2 \simeq G_F^2$ for $m_1 < m_3$
and $G_{\tau \mu}^2 \lesssim G_{\tau e}^2 \simeq G_F^2$ for $m_1 > m_3$
via the interaction with the matrix $X_\nu$.
 Therefore,
the Group-V and the Group-VII
could be tested at the Belle experiment
or the Belle-II experiment~\cite{Abe:2010gxa}.
 The other four groups can be tested
if some scalar particle is discovered at collider experiments.
 In this way,
our classification is useful
to discriminate mechanisms for generating Dirac neutrino masses
by testing not each model but each group of models.

\begin{acknowledgments}
 This work was supported, in part,
by Grant-in-Aid for Scientific Research No.~23104006~(SK)
and Grant H2020-MSCA-RISE-2014 No.~645722
(Non Minimal Higgs) (SK).
\end{acknowledgments}

\appendix
\section{Examples to close scalar lines in cases without dark matter}
\label{sec:app-A}

\begin{figure}[t]
 \begin{minipage}{0.45\hsize}
  \vspace*{1mm}
  \begin{center}
   \includegraphics[scale=0.7]{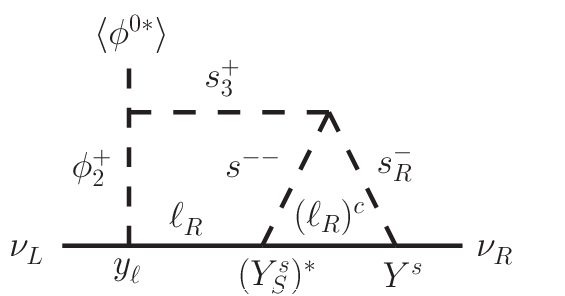}
   \vspace*{-3mm}
   \caption{
    An example to close scalar lines
    of the diagram in Fig.~\ref{fig:Dirac-3}.
   }
   \label{fig:loop-1}
  \end{center}
 \end{minipage}
\hspace*{5mm}
 \begin{minipage}{0.47\hsize}
  \begin{center}
   \includegraphics[scale=0.7]{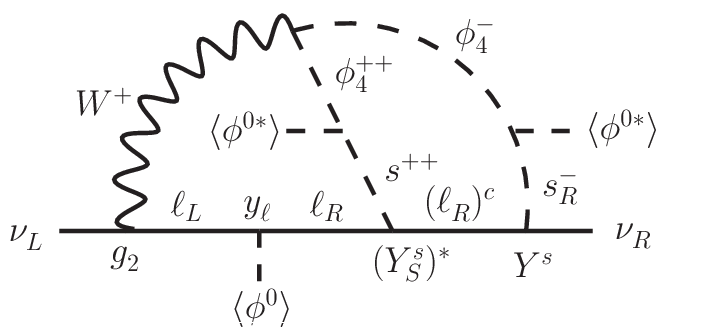}
   \vspace*{-13mm}
   \caption{
    An example to close scalar lines
    of the diagram in Fig.~\ref{fig:Dirac-4}.
   }
   \label{fig:loop-2}
  \end{center}
 \end{minipage}
\vspace*{5mm}
\end{figure}

 We show examples to close scalar lines
for Figs.~\ref{fig:NM-type}-\ref{fig:Dirac-6}
by using additional scalar fields in Table.~\ref{tab:particle-III}.
 Notice that these scalar fields
do not have Yukawa interactions.
 In Table~\ref{tab:loop-1},
we summarize scalar particles and relevant interactions
for each of Figs.~\ref{fig:NM-type}-\ref{fig:Dirac-6}.
 See also Figs.~\ref{fig:loop-1} and \ref{fig:loop-2}.

 For Fig.~\ref{fig:NM-type},
the example corresponds to
the model in Refs.~\cite{1loopDirac, Kanemura:2011jj}.
 The $Z_2^\prime$ symmetry is softly broken by $\mu^2$.
 For the other five figures listed in Table~\ref{tab:loop-1},
the parameter $\mu$ or $\mu^\prime$
softly breaks $Z_2^\prime$
whether the additional scalar is the $Z_2^\prime$-even or odd.
 Therefore,
we can confirm that
both of $\mu$ and $\mu^\prime$ are
necessary to close the scalar line
with the soft-breaking of $Z_2^\prime$.
 For Fig.~\ref{fig:nuTHDM-type},
which has only a scalar line,
explicit models can be found
in Refs.~\cite{Ref:nuTHDM-D, Davidson:2009ha}.

\begin{table}[t]
\renewcommand{\arraystretch}{1.2}
\begin{tabular}{c||c|c|c}
Scalar
 & $\SU(2)_L$
 & $\U(1)_Y$
 & L\#
\\[0.5mm]
\hline
\hline
$s_3^+$
 & ${\bf\underline{1}}$
 & $1$
 & $0$
\\[1mm]
\hline
$\Phi_3$
 & ${\bf\underline{2}}$
 & $\displaystyle\frac{1}{\,2\,}$
 & $-2$
\\[1mm]
\hline
\renewcommand{\arraystretch}{1}
$\Phi_4$
 & ${\bf\underline{2}}$
 & $\displaystyle\frac{3}{\,2\,}$
 & $-2$
\end{tabular}
\caption
{
 Examples of scalar fields
that can be used to close scalar lines
in Figs.~\ref{fig:NM-type}-\ref{fig:Dirac-6}
and Figs.~\ref{fig:Z2-1}-\ref{fig:GS-type}.
}
\label{tab:particle-III}
\vspace*{5mm}
\renewcommand{\arraystretch}{1}
\end{table}

\begin{table}[t]
\begin{tabular}{c||c|lll}
 {}
 & Scalar
 & \multicolumn{3}{|c}{Relevant interaction}
\\\hline\hline
 Fig.~\ref{fig:NM-type}
 & None
 & $\mu^2 [s_L^+ s_R^-]$
 & {}
 & {}
\\\hline
%
%
 Fig.~\ref{fig:Dirac-2}
 & $\Phi_3$
 & $\mu [\Phi_3^T \epsilon \Phi s_R^-]$ ,
 & $\mu^\prime [\Phi^\dagger \Delta \epsilon \Phi_3^\ast]$
 & {}
\\\hline
%
%
 Fig.~\ref{fig:Dirac-3}
 & $s_3^+$
 & $\mu [ s_R^- s^{++} s_3^- ]$ ,
 & $\mu^\prime [\Phi_2^\dagger \epsilon \Phi^\ast s_3^+ ]$
 & {}
\\\hline
%
%
 Fig.~\ref{fig:Dirac-4}
 & $\Phi_4$
 & $\mu [ \Phi^\dagger \Phi_4 s_R^- ]$ ,
 & $\mu^\prime [ \Phi_4^\dagger \epsilon \Phi^\ast s^{++} ]$
 & {}
\\\hline
%
%
 Fig.~\ref{fig:Dirac-5}
 & $s_3^+$
 & $\mu [ s^{0\ast} s_R^+ s_3^- ]$ ,
 & $\mu^\prime [\Phi_2^\dagger \epsilon \Phi^\ast s_3^+ ]$
 & {}
\\\hline
%
%
 Fig.~\ref{fig:Dirac-6}
 & $\Phi_3$
 & $\mu [\Phi_3^\dagger \epsilon \Phi^\ast s_R^+]$ ,
 & $\mu^\prime [\Phi^\dagger \Phi_3 (s^0)^\ast ]$
 & {}
\end{tabular}
\caption
{
 Examples of additional scalar fields and their interactions
to close scalar lines of Figs.~\ref{fig:NM-type}-\ref{fig:Dirac-6}.
}
\label{tab:loop-1}
\vspace*{5mm}
\end{table}

\section{Examples to close scalar lines in cases with dark matter}
\label{sec:app-B}

 We show examples to close scalar lines
for Figs.~\ref{fig:Z2-1}-\ref{fig:GS-type}
by using additional scalar fields in Table.~\ref{tab:particle-III}.
 In Table~\ref{tab:loop-2},
we summarize scalar particles and relevant interactions
for each of Figs.~\ref{fig:Z2-1}-\ref{fig:GS-type}.

 For Figs.~\ref{fig:Z2-1} and \ref{fig:GS-type},
scalar lines can be simply connected
without introducing additional scalar fields,
and the $Z_2^\prime$ symmetry is softly broken
by the parameter $\mu$.
 An explicit model for the structure in Fig.~\ref{fig:GS-type}
can be found in Ref.~\cite{Gu:2007ug}%
~(See also Ref.~\cite{Farzan:2012sa}).
 For Figs.~\ref{fig:Z2-2}-\ref{fig:Z2-7},
the parameter $\mu$ softly breaks $Z_2^\prime$
when we fix the $Z_2^\prime$ parity for
the additional scalar field as shown in Table~\ref{tab:loop-2}.
 Since the $Z_2^\prime$ parity for the scalar field
is fixed by $\lambda$
so that the term does not break $Z_2^\prime$,
the dimensionless coupling constant $\lambda$
is also necessary for the soft-breaking of $Z_2^\prime$.
 For Figs.~\ref{fig:Z2-8} and \ref{fig:Z2-10},
the product $\mu\mu^\prime$ softly breaks $Z_2^\prime$
independently on the $Z_2^\prime$ parity
of the additional scalar.
 For Fig.~\ref{fig:Z2-9},
the scalar lines can be closed
by introducing $(s_3^0)^\ast$~($\SU(2)_L$-singlet with $Y=0$)
in addition to $s_3^+$ and $\Phi_3$.
 Their lepton numbers are common and arbitrary.
 We additionally impose an unbroken $Z_2$ symmetry,
under which these three scalar fields have the odd parity.
 We see that the $Z_2^\prime$ symmetry
is softly broken by the product $\lambda \mu \mu^\prime$
independently on the $Z_2^\prime$ parities
of $(s_3^0)^\ast$, $s_3^+$, and $\Phi_3$.

 We obtain predictions
for the violation of the lepton universality
as shown in Table~\ref{tab:ltolnn}
by concentrating on the flavor structure.
 If we specify the scalar sector,
it is possible to perform further calculations.
 For example,
if scalar lines in Fig.~\ref{fig:Z2-8} of the Group-V
are closed by using $s_3^+$,
we have
\begin{eqnarray}
(G_{\ell\ellp}^X)^2
=
 \left( \frac{1}{ 16\pi^2 } \right)^2
 \frac{
       (m_\text{D}^{} m_\text{D}^\dagger)_{\ell\ell}
       (m_\text{D}^{} m_\text{D}^\dagger)_{\ellp\ellp}
      }
      { 8\, \Lambda^4 C_\text{loop}^4 m_\ell^2 m_{\ellp}^2 } , \quad
%
%
C_\text{loop}
=
 \left( \frac{1}{16\pi^2} \right)^2
 \frac{ \mu\mu^\prime }{ \Lambda^2 }
\end{eqnarray}
 By taking $(G^X_{e\mu}/G_F)^2 = 10^{-3}$
with $m_\text{D} = 0.1\,\eV$ for example,
we see $\mu\mu^\prime/\Lambda = {\mathcal O}(10^{-2})\,\GeV$.

\begin{table}[t]
\begin{tabular}{c||c|lll}
 {}
 & Scalar
 & \multicolumn{3}{|c}{Relevant interaction}
\\\hline\hline
 Fig.~\ref{fig:Z2-1}
 & None
 & $\mu [s_L^+ s_2^- (s_2^0)^\ast]$
 & {}
 & {}
\\\hline
%
%
 Fig.~\ref{fig:Z2-2}
 & $\Phi_3$ ($Z_2^\prime$-odd)
 & $\mu [\Phi^T \Delta \epsilon \Phi_3^\ast]$ ,
 & $\lambda [\Phi_3^T \epsilon \Phi s_2^- (s_2^0)^\ast]$
 & {}
\\\hline
%
%
 Fig.~\ref{fig:Z2-3}
 & $\Phi_3$ ($Z_2^\prime$-even)
 & $\mu [\Phi_3^T \epsilon \Phi s_R^-]$ ,
 & $\lambda [(\Phi^\dagger \eta)(\Phi_3^\dagger \eta)]$
 & {}
\\\hline
%
%
 Fig.~\ref{fig:Z2-4}
 & $s_3^+$ ($Z_2^\prime$-odd)
 & $\mu [\Phi_2^\dagger \epsilon \Phi^\ast s_3^+]$ ,
 & $\lambda [s_3^- s_R^- s_2^+ s_2^+]$
 & {}
\\\hline
%
%
 Fig.~\ref{fig:Z2-5}
 & $\Phi_4$ ($Z_2^\prime$-even)
 & $\mu [\Phi^\dagger \Phi_4 s_R^-]$ ,
 & $\lambda [\Phi_4^\dagger \epsilon \Phi^\ast s_2^+ s_2^+]$
 & {}
\\\hline
%
%
 Fig.~\ref{fig:Z2-6}
 & $s_3^+$ ($Z_2^\prime$-odd)
 & $\mu [\Phi_2^\dagger \epsilon \Phi^\ast s_3^+]$ ,
 & $\lambda [(s_2^0)^\ast (s_2^0)^\ast s_R^+ s_3^-]$
 & {}
\\\hline
%
%
 Fig.~\ref{fig:Z2-7}
 &
 $\Phi_3$
 ($Z_2^\prime$-even)
 &
 $\mu [\Phi_3^\dagger \epsilon \Phi^\ast s_R^+]$ ,
 &
 $\lambda [\Phi^\dagger \Phi_3 (s_2^0)^\ast (s_2^0)^\ast]$
 & {}
\\\hline
%
%
 Fig.~\ref{fig:Z2-8}
 & $s_3^+$
 & $\mu [ \Phi_2^\dagger \epsilon \Phi^\ast s_3^+ ]$ ,
 & $\mu^\prime [ s_3^- s_2^+ (s_2^0)^\ast ]$
 & {}
\\\hline
%
%
 Fig.~\ref{fig:Z2-9}
 & $(s_3^0)^\ast$ , $s_3^+$ , $\Phi_3$
 & $\mu [\Phi^\dagger \Phi_3 s_3^0]$ ,
 & $\mu^\prime [\Phi_3^\dagger \epsilon \Phi^\ast s_3^+]$ ,
 & $\lambda [(s_3^0)^\ast s_3^- (s_2^0)^\ast s_2^+]$
\\[-3mm]
 {}
 & ($Z_2$-odd, unbroken)
 & {}
 & {}
 & {}
\\\hline
%
%
 Fig.~\ref{fig:Z2-10}
 & $\Phi_4$
 & $\mu [\Phi_4^\dagger \eta s_2^+]$ ,
 & $\mu^\prime [\Phi^\dagger \Phi_4 s_R^-]$
 & {}
\\\hline
%
%
 Fig.~\ref{fig:GS-type}
 & None
 & $\mu[\Phi^\dagger \eta (s_2^0)^\ast]$
 & {}
 & {}
\end{tabular}
\caption
{
 Examples of additional scalar fields and their interactions
to close scalar lines of Figs.~\ref{fig:Z2-1}-\ref{fig:GS-type}.
 For Fig.~\ref{fig:Z2-9},
a common L\# is assigned to these additional scalar fields,
where $s_3^0$ is a gauge singlet field.
 Then,
an unbroken $Z_2$ symmetry is imposed
such that these scalar fields have the odd parity.
}
\label{tab:loop-2}
\vspace*{5mm}
\end{table}

\end{document}